\documentclass[twocolumn,aps,prl,showpacs,groupedaddress]{revtex4}
\usepackage{amssymb}
\usepackage{epsfig,amssymb,amsmath}

\newcommand{\ket}[1]{|#1\rangle}

\newcommand{\eq}{\begin{equation}}
\newcommand{\fine}{\end{equation}}

\begin{document}

\title{Coherent scattering of a Multiphoton Quantum Superposition by a Mirror-BEC}
\author{Francesco De Martini$^{1,2}$,Fabio Sciarrino$^{1}$, Chiara Vitelli$%
^{1}$}
\author{Francesco S. Cataliotti$^{3}$}

\begin{abstract}
We present the proposition of an experiment in which the multiphoton quantum
superposition consisting of $\mathcal{N}\approx 10^{5}$ particles generated
by a quantum-injected optical parametric amplifier (QI-OPA), seeded by a
single-photon belonging to an EPR entangled pair, is made to interact with a
\textit{Mirror-BEC} shaped as a Bragg interference structure. The overall
process will realize a Macroscopic Quantum Superposition (MQS) involving a
microscopic single-photon state of polarization entangled with the coherent
macroscopic transfer of momentum to the BEC structure, acting in space-like
separated distant places.
\end{abstract}

\affiliation{(1) Dipartimento di Fisica, Universit\`{a} di Roma "La Sapienza", Piazzale
Aldo Moro 2, I-00185 Roma, Italy}
\affiliation{(2) Accademia Nazionale dei Lincei, Via della Lungara 10, I-00165 Roma,
Italy.}
\affiliation{(3) Dipartimento di Energetica and LENS, Universit\`{a} di Firenze, via N.
Carrara 1, I-50019 Sesto F.no (FI), Italy}
\maketitle

In recent years a great deal of interest has been attracted by the
ambitious problem of creating a Macroscopic Quantum Superposition
(MQS)\ of a massive object by an entangled opto-mechanical
interaction of a tiny mirror with a single photon within a
Michelson interferometer \cite{Marshall03, Bohm06, Giga06, Arci06,
Klec06}, then realizing a well known 1935 argument by Erwin
Schr\"{o}dinger \cite{Schroe35}. The present work is aimed at a
similar scope but is not concerned with interferometers nor with
solid mirrors. It rather exploits the process of
nonresonant scattering by a properly shaped Bose--Einstein condensate (BEC)%
\cite{Inguscio} of an externally generated multi-particle quantum photon
state, a \textquotedblright macro-state\textquotedblright\ $\left\vert \Phi
\right\rangle $, in order to create a joint atom-photon macro-state
entangled by momentum conservation. Light scattering from BEC structures\
has been used so far to enhance their non--linear macroscopic properties in
super--radiance experiments \cite{Inouye99}, to show the possibility of
matter wave amplification \cite{Kozuma99} and non-linear wave mixing \cite%
{Deng99}. In the present work we intend to discuss the linear \textit{%
coherent scattering, i.e. }the\textit{\ reflection} by a multilayered BEC of
a large assembly of nearly monochromatic photons generated by a high-gain
\textquotedblright quantum-injected\textquotedblright\ Optical Parametric
Amplifier (QI-OPA) in a Einstein-Podolsky-Rosen (EPR) configuration \cite%
{DeMa98,DeMa05}. Very recently it was demonstrated experimentally
that the optical macrostate $\left\vert \Phi \right\rangle $
generated by the QI-OPA can indeed be entangled with, i.e.
non-separable from, a\ far apart single photon state belonging to
the injected EPR\ pair \cite{DeMa08}, thus resulting highly
resilient to the decoherence due to losses \cite{DeMa09}. By the
present work this condition\ will be extended to the mechanical
motion of an atomic assembly by making the photonic macrostate
$\left\vert \Phi \right\rangle $ to exchange linear momentum with
a high reflectivity BEC\ optical mirror, here referred to as a
\textit{\textquotedblright Mirror-BEC\textquotedblright\ }(M-BEC).
This can be a novel and viable
alternative to the realization of an entangled MQS of a massive object.%
\newline
\begin{figure}[t]
\includegraphics[width=8 cm]{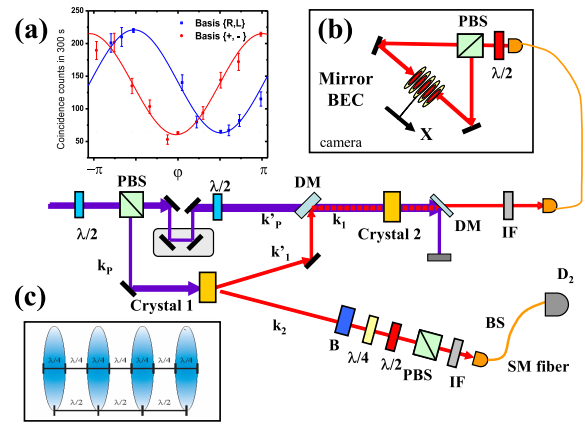}
\caption{Layout of the QI-OPA + Mirror BEC experimental apparatus.
The upper left (a)\ Inset shows the interference\ patterns
detected at the output of the PBS\ shown in the\ (b)-Inset for two
different measurement basis $\left\{ +,-\right\} $ and $\left\{
L,R\right\} $. Alternating slabs of condensate and vacuum are
shown in the lower left (c)\ Inset. A more detailed account of the
Inset (b) is given in Fig.3.} \label{schema}
\end{figure}
The layout of the experiment, Figure \ref{schema}, shows an EPR optical
parametric amplifier, provided by Crystal 1, of a polarization entangled ($%
\pi -$en$\tan $gled) pair of photons launched towards two distant
measurement stations, here referred to as \textit{Alice} (A)\ and \textit{Bob%
} (B) \cite{DeMa08,Naga07}. One of the EPR\ photons emitted
towards the Bob's site is injected into the QIOPA which generates
a corresponding macrostate $\left\vert \Phi \right\rangle $. The
device operates in the collinear regime and amplifies with a large
\textquotedblright gain\textquotedblright\ any injected single
photon in a quantum superposition, i.e. a \textit{qubit
}$\left\vert \varphi \right\rangle _{k1}$ into a large number of
photons, $\mathcal{N}\approx 10^{5}$, associated with a
corresponding macro-qubit $\left\vert \Phi ^{\varphi
}\right\rangle _{k1}$. These macrostates then drive the mechanical
motion of the Mirror-BEC. Since these states are found to be
entangled with the far part single-photon emitted over the mode
$\mathbf{k}_{2}$ and detected by Alice, the same entanglement
property is then transferred to the position-macrostate of the
optically-driven Mirror-BEC ((b)-inset). The optical part of the
apparatus is the working QI-OPA device recently reported by
\cite{DeMa08,Naga07} to which the reader is referred.

\textsl{Micro - Macro entangled light. }As shown in Fig.1, the main UV beam
is split in two beams and excites two nonlinear (NL) crystals cut for type
II phase-matching. Crystal 1 is the spontaneous parametric down-conversion
source of entangled photon couples of wl $\lambda ^{\prime }=2\lambda
_{P}^{\prime }$, emitted over the modes $\mathbf{k}_{i}$ ($i=1^{\prime },2$)
in the entangled \textit{singlet} state $\left\vert \Psi ^{-}\right\rangle
_{k1^{\prime },k2}$= $2^{-1/2}\left( \left\vert H\right\rangle _{k1^{\prime
}}\left\vert V\right\rangle _{k2}-\left\vert V\right\rangle _{k1^{\prime
}}\left\vert H\right\rangle _{k2}\right) $,$\ $where $H(V)$ labels the
single photon state horizontally (vertically) polarized $.$ The photon
associated with the mode $\mathbf{k}_{2}$ (the \textit{trigger} mode) is
coupled to a single mode (SM)\ fiber and filtered by a set of $\pi -$%
analysing optical devices, namely a Babinet compensator (B) a
$\lambda ^{\prime }/2$ $+$ $\lambda ^{\prime }/4$ waveplate set, a
PBS and an interference filter (IF) with a transmission linewidth
$\Delta \lambda ^{\prime }$. At last the \textit{trigger} photon
excites, at the
Alice's site, the single photon detector $D_{2}^{A}$ delivering the \textit{%
trigger }signal adopted to establish the overall quantum correlations. By a
Dichroic Mirror (DM) the single photon created over the mode $\mathbf{k}%
_{1}^{\prime }$ is made to merge into the mode $\mathbf{k}_{1}$
together with the UV  beam associated with mode
$\mathbf{k}_{P}^{\prime }$, and then injected into the NL Crystal
2 where it stimulates the emission of many photon pairs over the
two polarization output modes associated with the spatial mode
$\mathbf{k}_{1}$. \ The injected qubit is prepared in the state
$\left\vert \varphi \right\rangle _{k1}=2^{-1/2}(\left\vert
H\right\rangle _{k1}+e^{i\varphi }\left\vert V\right\rangle
_{k1})$ by measuring the photon on mode $\mathbf{k}_{2}$ in an
appropriate polarization basis.

When a single photon qubit $\left\vert \pm \right\rangle
=2^{-1/2}(\left\vert H\right\rangle \pm \left\vert V\right\rangle
)$ is injected on mode $\mathbf{k}_{1}$, the amplified output
state is expressed as $\left\vert \Phi ^{\pm }\right\rangle
=\sum\limits_{i,j=0}^{\infty }\gamma _{ij}\ket{(2i+1)\pm
}\ket{(2j)\mp}$ and $\gamma
_{ij}\equiv \sqrt{(1+2i)!(2j)!}(i!j!)^{-1}C^{-2}(-\frac{\Gamma }{2})^{j}%
\frac{\Gamma }{2}^{i}$, $C\equiv \cosh g$, $\Gamma \equiv \tanh g$, being $g$%
\ $=$ $\chi t$ the NL\ gain. There $\ket{p+}\ket{q-}$ stands for a
state with $p$ photons with polarization $\overrightarrow{\pi
}_{+}$
and $q$ photons with $\overrightarrow{\pi }_{-}$. The macro-states $%
\left\vert \Phi ^{+}\right\rangle $, $\left\vert \Phi ^{-}\right\rangle $
are orthonormal, i.e. $\left\langle \,^{i}\Phi |\Phi ^{j}\right\rangle
=\delta _{ij}$. The overall entangled  $%
\left\vert \Sigma \right\rangle _{k1,k2}=$\ $2^{-1/2}\left( \left\vert \Phi
^{+}\right\rangle _{k1}\left\vert +\right\rangle _{k2}-\left\vert \Phi
^{-}\right\rangle _{k1}\left\vert -\right\rangle _{k2}\right) $ keeps its
\textit{singlet} character in the multi-particle regime between two distant objects: the \textit{microscopic,}
i.e. single particle system expressed by the \textit{trigger} state (mode $%
\mathbf{k}_{2}$) and the \textit{macroscopic,} multiparticle system ($%
\mathbf{k}_{1}$) \cite{Schl01}. At the output of crystal $2$ the
beam with wl $\lambda ^{\prime }$ is spatially separated by the
pump UV beam by a DM and an IF with bandwidth $0.75nm$ and finally
coupled to a single mode fiber. Two counteracting optical beams
associated with the
macrostates $\left\vert \Phi ^{\pm }\right\rangle $ for a total of $\mathcal{%
N}\simeq 3\overline{m}\approx 1.2\times 10^{5}$ are spatially separated by a
$PBS$\ and focused on the opposed sides of a cigar-shaped M-BEC.\newline
\begin{figure}[h]
\includegraphics[width=0.5\textwidth]{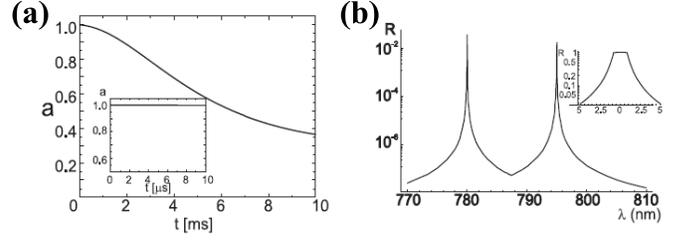}
\caption{(\textbf{a}) Normalized amplitude $a=I(t)/I(0)$, where
$I(t)$ is the intensity of the reflected beam at time $t$ after
the interaction, of the beam reflected by the \textit{Mirror BEC}
in the proposed experimental conditions. In the inset it
is shown the normalized amplitude in the first 10 $\protect\mu $s. (\textbf{b}%
) Reflectivity of a patterned BEC as a function of wavelength. In
the inset is shown the reflectivity around resonance}
\label{carlevo}
\end{figure}
Let us analyze the output field over the polarization modes $\overrightarrow{%
\pi }_{\pm }$ i.e. by adopting the measurement basis $\left\{ +,-\right\} $
realized by qubit $\left\vert \varphi \right\rangle $. The ensemble average
photon number $N_{\pm }$ emitted over $\mathbf{k}_{1}$ depends on the
injected phase $\varphi $: $N_{\pm }(\varphi )\;$= $\left\langle ^{\varphi
}\Phi \right\vert \widehat{N}_{\pm }(\varphi )\left\vert \Phi ^{\varphi
}\right\rangle $ = $\overline{m}+(\overline{m}+\frac{1}{2})(1\pm \cos
\varphi )$ with $\overline{m}=\sinh ^{2}g$, the average value of the number
of \ \textit{squeezed vacuum}\ photons emitted by OPA\ in absence of
injection \cite{DeMa98}. The number difference: $N(\varphi )\;$= $%
[N_{+}(\varphi )-N_{-}(\varphi )]$ = $[(2\overline{m}+1)\cos \varphi ]\;$is
expressed by an interference fringe pattern as function of the phase $%
\varphi $ of both the quantum-injected qubit and the\ EPR
correlated trigger\ qubit (Figure 1 ((a)-Inset)).

\emph{BEC mirror via Bragg reflection}. Let us now describe the structure of
the M-BEC and its interaction with light. The dynamics of a BEC loaded in a
trap formed by a cylindrically symmetric harmonic potential (either an
optical trap or a magnetic trap) with an optical standing - wave (SW)
aligned along the symmetry axis, may be described by the Gross-Pitaevskii
equation \cite{GP}. If the trap is very elongated the ground state consists
of an array of disks spaced by half the wavelength $\lambda $ with a
longitudinal size $R_{l}\propto \lambda /s^{-1/4}$ with $s$ being an
adimensional parameter describing the height of the optical lattice in terms
of the recoil energy $E_{R}=h^{2}/2m\lambda ^{2}$. The transverse size $%
R_{\perp }$ is dictated by the strength of the magnetic trap and
by the number of atoms in the condensate $N_{at}$ \cite{Pedri01}.
The number $N_{D}$ of the disks is also fixed by the strength of
the magnetic trap in the longitudinal direction and by $N_{at}$.
Typical numbers are $N_{D}\sim 200$, $N_{at}\sim 10^{6}$ with
$R_{\perp }\sim 10\mu $m \cite{Morsch06}. By choosing $s$ it is
then possible to prepare an array of disks with a longitudinal
size of $R_{l}=\frac{\lambda }{4}$ spaced by $\frac{\lambda }{2}
$: Figure 1 (Inset (c)).

Releasing the condensate from the combined trap, the spatial
periodic structure is initially preserved as long as the spreading
disks do not start to overlap and interfere, and eventually leads
for longer expansion times to a structure that reflects the
momentum distribution of the condensate. Both regimes are
fundamental to our proposal. If we expose the expanding condensate
aligned along the symmetry axis of the harmonic trap to an optical
beam with frequency $\omega -$ $\omega _{0}=\Delta \ $, largely
detuned from the atomic resonance $\omega _{0}$, the dominant
scattering mechanism is Rayleigh scattering \cite{Inouye99}. The
dynamic evolution of the system in this regime is described by the
1-D CARL-BEC i.e. Gross-Pitaevskii, model generalized to include
the self-consistent evolution of the scattered radiation amplitude
\cite{Boni04,DeSarlo05,CARL}:

{\small
\begin{eqnarray}
i\frac{\partial \Psi }{\partial t} &=&-\frac{\hbar ^{2}}{2m}\frac{\partial
^{2}\Psi }{\partial x^{2}}+ig\left\{ b^{\ast }e^{i(2kx-\delta t)}-\mathrm{%
c.c.}\right\} \Psi  \label{psi} \\
\frac{da}{dt} &=&gN\int dx|\Psi |^{2}e^{i(2kx-\delta t)}-\kappa a.  \label{a}
\end{eqnarray}%
} \noindent

with $b=(\epsilon _{0}V/2\hbar \omega _{s})^{1/2}E_{s}$ the
dimensionless electric field amplitude of the reflected beam with
frequency $\omega _{s}$, $g=(\Omega /2\Delta )(\omega d^{2}/2\hbar
\epsilon _{0}V)^{1/2}$ the coupling constant, and $\Omega $ the
Rabi frequency modulation of the optical beam,
$d=\hat{\epsilon}\cdot \vec{d}$ is the electric dipole moment
of the atom along the polarization direction $\vec{\epsilon}$ of the laser, $%
V$ is the volume of the condensate, $N$ is the total number of
atoms in the condensate and $\delta =\omega -\omega _{s}$
\cite{Equat}. Let us focus on the amplitude of the reflected beam.
We have integrated numerically the Equations (\ref{psi},\ref{a})
with the experimental parameters (given above) of a typical
condensate and with the optical parameters for the output of the
QI-OPA. As shown by Figure 2-\textbf{(a)}, the normalized
amplitude $a$ of the reflected optical beam drops as the matter
wave grating is deteriorated by the interaction with the light
beam. However, during the time duration of a QI-OPA pulse
(typically 1 ps) no significant reduction is observed (inset of Figure \ref%
{carlevo}-(\textbf{a})).

The above result leading to the dependence on $\omega $ of the M-BEC\
reflectivity has been found consistent with a less sophisticated model of
the process based on a classical model for a Bragg mirror. The reflectivity
of this one composed by $2N_{D}$ alternating layers with refraction index $%
n_{B}\sim (1+\epsilon )$ is: $R=\left( \frac{n_{B}^{2N_{D}}-1}{%
n_{B}^{2N_{D}}+1}\right) ^{2}\sim N_{D}^{2}\epsilon ^{2}$.%
For a 2--level atom $\epsilon =\frac{3\pi }{2}{\mathcal{M}}4\frac{\Gamma }{%
\Delta }\frac{1}{1+\left( 2\frac{\Delta }{\Gamma }\right) ^{2}}$
being ${\mathcal{M}}=\lambdabar ^{3}\frac{N}{V}$ is the rescaled density, $%
\Gamma $ the atomic linewidth and $\Delta $ the detuning from resonance. In
a rubidium BEC:$\ \Gamma \simeq $ 6 MHz and typical densities are $\frac{N}{V%
}=10^{14}$cm$^{-3}$. Combining all the previous equations and
assumptions we obtain the graph Figure 2-(\textbf{b}). The inset
of this Figure shows that around the atomic resonance with a
bandwidth $\Delta \nu _{a}\simeq 8GHz$ the reflectivity of the
patterned BEC is essentially unity \cite{DetailBW}. This
bandwidth, three orders of magnitude larger than in other
proposals \cite{artoni} is instrumental to the proposed experiment
sketched in the right inset of Fig.1. A first estimate of the
experimental parameters of the QIOPA system results in a NL gain
$g=6-7$ corresponding to a number of generated photons $\sim
10^{5}-10^{6}$. Since the spectral width of the QI-OPA generated
beams is $\Delta \lambda \sim 0.75nm$, corresponding to a
linewidth $\Delta \nu \sim 350GHz$, about the 3\% of the incoming
photon
beams will be totally reflected by M-BEC. 
This will corresponds to a number of active photons $N_{\pm }^{\prime
}(\varphi )$ = $(\Delta \nu _{a}/\Delta \nu )\times N_{\pm }(\varphi )$ in
the range $(10^{3}-10^{4})$. 
The ratio between the linewidths relative to the absorption (few MHz) and to
the reflection process is $\simeq $ 0.1\% hence the mean number of absorbed
photons is about one per pulse, it follows that the excitation of atoms can
be considered negligible during the interaction process. At last, in the
future we plan to use a different laser source with a longer pulse duration
followed by a periodically-poled crystal amplifier. In such a way, it would
be possible to obtain a high NL gain value and radiation fields with a
bandwidth of $\sim 10GHz$.
\begin{figure}[b]
\includegraphics[width=0.5\textwidth]{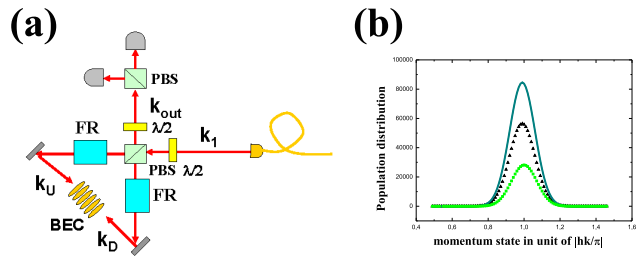}%
\caption{\textbf{a}: Interaction between the amplified field and
the Bragg Mirror BEC. The Faraday Rotators (FR) allows the
recombination the reflected field on mode $k_{out}$. Indeed the
field with polarization $\protect\pi _{H}
$ transmitted by the PBS on mode $k_{U}$, passing through the first FR at $%
22.5%
{{}^\circ}%
$, becomes $\protect\pi _{+}$ polarized. After the reflection by
the Bragg Mirror its polarization is again rotated and becomes
$\protect\pi _{V}$: the field exits on mode $k_{out}$. A similar
argumentation holds for the field polarized $\protect\pi _{V}$.
\textbf{b:} comparison between the expected populations of the
first order momentum peaks (around $1\%$ of total atoms) before
and after the interaction. The dotted (triangles) line represents
the profile of the $\pm2\hbar k$ momentum state before interaction
with the QIOPA pulse, the continuous line and dotted (circles)
line represent the $+2\hbar k$ and $-2\hbar k$ momentum state
profiles after interaction, respectively. Both heating and
collisional effects have been neglected (see text).}
\label{fig:measurement}
\end{figure}

\textsl{Measurement of nonlocal correlations.} In order to observe the
recoil effects of the \emph{M-BEC}, the condensate has to be released from
the optical lattice that shapes it. This can easily be done by shutting off
the lasers light that provides the SW. Typical expansion velocities for a
BEC are of the order of 1 nm/$\mu $s which leaves at least 50 $\mu $s before
the pattern gets significantly spoiled. Let's investigate  the different
evolution of the BEC motions relative to the impinging field. Consider the
case in which the multiphoton field is prepared in the state $|\Phi
^{+}\rangle $ (or, alternatively: $|\Phi ^{-}\rangle $) on the spatial mode $%
\mathbf{k}_{1}$. The state can be written as: $|\Phi ^{+}\rangle _{1}=|\phi
^{+}\rangle _{1}|\xi ^{-}\rangle _{1}$ ($|\Phi ^{-}\rangle _{1}=|\phi
^{-}\rangle _{1}|\xi ^{+}\rangle _{1}$), where $|\phi ^{+}\rangle $ ($|\phi
^{-}\rangle $) is the wavefunction contribution with polarization $\vec{\pi}%
_{+}$($\vec{\pi}_{-}$) and $|\xi \rangle ^{-}$($|\xi \rangle
^{+}$) is the contribution with polarization
$\vec{\pi}_{-}$($\vec{\pi}_{+}$) .  The number of photons
associated to $|\phi \rangle $ is dominant over the one associated
to $|\xi \rangle $, as said. The multiphoton state $|\Phi
^{+}\rangle _{1}$ ($|\Phi ^{-}\rangle _{1}$) is sent, through a
single mode fiber, toward the BEC condensate. There, a $\lambda
/2$ waveplate and a
polarizing beam splitter direct the two polarization components $\vec{\pi}%
_{+}$ and $\vec{\pi _{-}}$ over the two spatial modes $\mathbf{k_{U}}$ and $%
\mathbf{k_{D}}$: the macrostate $|\Phi ^{+}\rangle _{1}$ ($|\Phi
^{-}\rangle _{1}$) evolves into $|\phi ^{+}\rangle _{U}|\xi
^{-}\rangle _{D}$ ($|\xi ^{+}\rangle _{U}|\phi ^{-}\rangle _{D}$):
Figure 3-\textbf{(a)}. Then, the two counterpropagating fields are
focused on the opposed sides of the cigar-shaped M-BEC. Thanks to
the large reflectivity of the Bragg structure, an efficient
coupling is achieved between the multiphoton fields and the
atomic cloud. 
While the back-scattered light pulse changes direction of propagation $%
(U\Rightarrow D,D\Rightarrow U)$, the M-BEC acquires a momentum
kick in the opposite direction to the major photons contribution
$|\phi ^{+}\rangle _{D}$ ($|\phi ^{-}\rangle _{D}$). Hence, after
the interaction the overall light-matter state can be written as:
$|\phi ^{+}\rangle _{D}|\xi ^{-}\rangle _{U}|\Psi ^{U}\rangle
_{bec}$ ( $|\xi ^{+}\rangle _{D}|\phi
^{-}\rangle _{U}|\Psi ^{D}\rangle _{bec}$) where $|\Psi ^{U}\rangle _{bec}$ (%
$|\Psi ^{D}\rangle _{bec}$) stands for the BEC that recoils in the
direction $\mathbf{k_{U}}$ ($\mathbf{k_{D}}$). Two Faraday
rotators inserted into the Sagnac-like interferometer allow the
recombination of both the reflected fields $\{|\phi ^{+}\rangle
_{D},|\xi ^{-}\rangle _{U}\}$ ($\{|\xi ^{+}\rangle _{D},|\phi
^{-}\rangle _{U}\}$) on the mode $\mathbf{k_{OUT}}$ leading to the
output state $|\Phi ^{+}\rangle _{OUT}|\Psi ^{U}\rangle _{bec}
$ ($|\Phi ^{-}\rangle _{OUT}|\Psi ^{D}\rangle _{bec}$). 

In the analysis above, $|\Psi \rangle _{bec}$ represents the state
of the interacting portion rather than of the overall BEC system.
Indeed the interaction with the QI-OPA pulse does not shift the
BEC as a whole but only the atoms which get the momentum kick by
the impinging photons. This mechanism of momentum transfer between
light and atoms has been experimentally investigated in different
works \cite{Gret05, Ande06}. For a
generic input\ photon macro-state $\left\vert \Phi ^{\varphi }\right\rangle $%
, the resulting momentum exchange due to any elementary interaction will
cause the spatial \textquotedblright displacement\textquotedblright\ of the
M-BEC 
depending on the phase $\varphi $ encoded in the far apart Alice's qubit
\cite{displacement}. The velocity acquired by the M-BEC is $\frac{N^{\prime }%
}{N_{at}}v_{r}$, where $v_{r}$ is the condensate recoil velocity of rubidium
(in this case around 5 mm/s). This is visible during expansion where, due to
the quantized nature of the momentum transfer, it appears as a transfer of
atoms from the lower momentum state to a higher momentum state. Precisely,
the normal momentum distribution of the M-BEC is made of sharp peaks
centered around zero. Because of the photon-atom collisions a large number
of atoms, for a total $N_{\pm }^{\prime }(\varphi )$, will be transferred
from the generic momentum state $n2h/\lambda $ to the successive state $%
(n\pm 1)2h/\lambda $.
The momentum state distribution then becomes asymmetric as
reported in Figure 3-\textbf{(b)}. There we show the result of a
numerical simulation of the population transfer from the zero
momentum to the fist order momentum state, and we compare the
population distributions after the interaction relative to the
$-2\hbar k$ and $+2 \hbar k$ momentum states.
The lifetime of the macroscopic quantum superposition is limited
by the decoherence between the original and the recoiled atomic wavepackets. In Ref. \cite{DeSarlo05} the decoherence rate for a matter wave grating formed due to the CARL effect has been found experimentally to be $6.4(9) ms^{-1}$. This should leave enough time to perform
a measurement of nonlocal correlations.
A preliminary
experimental test on a sample M-BEC with a classical light beam
indeed showed a marked shift of the atomic momentum distribution
due to light
collision and a reflectivity in the range $0.5-0.9$ . 
As a further improvement to this scenario, in the experiment only the
largest, most efficient optical pulses could be singled out by a ultra-fast
electro-optical switch placed at the output of the QI-OPA \cite{Giac08}.

In conclusion, the entanglement structure of $\left\vert \Sigma
\right\rangle _{k1,k2}$ will imply the coherent displacement of the M-BEC
system, depending on the phase $\varphi $ of
the single photon-measured by the far apart Alice's apparatus. The
correlation measurements could be carried out by detecting the reflected
light in different polarization basis and by observing the momentum
distribution of the atomic cloud. In addition, the reflection process effect
could be repeated several times by one or several external mirrors
reflecting back the optical beams to the M-BEC after the first interaction,
leading an optical \textit{cavity} structure, e.g. a Fabry-Perot
interferometer, by which the BEC displacing effect could be enhanced by a
\textquotedblright quality factor\textquotedblright\ $Q>>1$. 
We acknowledge interesting discussions with C. Fort, L. Fallani,
M. Inguscio, F.D.M. also acknowledges early discussions with
Markus Aspelmeyer leading to the basic idea of the present
proposal. Work supported by MIUR (N. PRIN 06), FP7 CHIMONO.


\begin{thebibliography}{99}
\bibitem{Equat} The last term in right-hand side of Eq.(\ref{psi})
represents the self-consistent optical wave grating while the first term in
the right-hand side of Eq.(\ref{a}) represents the self-consistent
matter-wave grating. Eq.(\ref{a}) has been written in the \textquotedblleft
mean-field\textquotedblright\ limit, which models the propagation effects by
a damping term where $\kappa \approx c/2L$ and $L$ is the length of the
condensate.
\bibitem{DetailBW} A more careful model should take into account a quasi
sinusoidal modulation of the refractive index due to the atomic density
distribution in the Thomas--Fermi approximation.
\end{thebibliography}
\end{document}